\documentclass[conference]{IEEEtran}
\IEEEoverridecommandlockouts

\usepackage{cite}
\usepackage{amsmath,amssymb,amsfonts}
\usepackage{algorithmic}
\usepackage{graphicx}
\usepackage{textcomp}
\usepackage[dvipsnames]{xcolor}
\usepackage{listings}
\usepackage[hidelinks]{hyperref}

\lstdefinestyle{cppstyle}{
    language=C++,
    basicstyle=\ttfamily\footnotesize,
    keywordstyle=\bfseries,
    commentstyle=\color{gray!70}\itshape,
    showstringspaces=false,
    numbers=left,
    numberstyle=\tiny\color{gray},
    numbersep=6pt,
    xleftmargin=14pt,
    aboveskip=4pt,
    belowskip=4pt,
    captionpos=b,
    morekeywords={nullptr,override}
}
\lstdefinestyle{jsonstyle}{
    basicstyle=\ttfamily\scriptsize,
    showstringspaces=false,
    breaklines=true,
    breakatwhitespace=true,
    columns=fullflexible,
    xleftmargin=4pt,
    aboveskip=4pt,
    belowskip=4pt,
    captionpos=b,
    literate={\{}{{{\color{black}\{}}}1
             {\}}{{{\color{black}\}}}}1
             {[}{{{\color{black}[}}}1
             {]}{{{\color{black}]}}}1
}
\def\BibTeX{{\rm B\kern-.05em{\sc i\kern-.025em b}\kern-.08em
    T\kern-.1667em\lower.7ex\hbox{E}\kern-.125emX}}

\begin{document}

\title{ATLAS: Multi-View Code Representation Tool for C and C++ Source Programs}

\author{
\IEEEauthorblockN{Jaid Monwar Chowdhury\textsuperscript{\textdagger*}, Ahmad Farhan Shahriar Chowdhury\textsuperscript{\textdagger\S}, Humayra Binte Monwar\textsuperscript{\textdaggerdbl\S}, Mahmuda Naznin\textsuperscript{\textdagger}}
\IEEEauthorblockA{\textsuperscript{*}\textit{Dept. of Computer Science, University of Texas at Arlington, Arlington, TX, USA}\\
jxc3648@mavs.uta.edu\\
\textsuperscript{\textdagger}\textit{Dept. of Computer Science and Engineering, Bangladesh University of Engineering and Technology, Dhaka, Bangladesh}\\
\{1905081@ugrad.cse.buet.ac.bd, mahmudanaznin@cse.buet.ac.bd\}\\
\textsuperscript{\textdaggerdbl}\textit{Dept. of Electrical and Electronic Engineering, Bangladesh University of Engineering and Technology, Dhaka, Bangladesh}\\
2006159@eee.buet.ac.bd\\
\textsuperscript{\S}Equal contribution}
}

\maketitle

\begin{abstract}
Multi-view code graphs that align abstract syntax trees, control flow graphs, and data flow graphs are now central to machine-learning models for software engineering. For C and C++, no single tool produces these aligned views without a complete build. We present ATLAS, a command-line tool that takes one or more C or C++ source files and emits an AST, a source-level inter-procedural CFG, a reaching-definition DFG, or any combination. The output is available as JSON, DOT, or PNG. ATLAS runs directly on source code and accepts partial input such as a project with missing headers, so it needs no compilation or build database. All views share one node namespace, so a downstream consumer can recover any single view by filtering edges. Command-line flags select views, collapse variables, and blacklist node categories to resize the emitted graph. On the TheAlgorithms corpora, ATLAS produces a correct CFG for $96.80\%$ of C files and $91.67\%$ of C++ files, including files that do not compile. It already serves as the CFG front-end of an LLM-based unit-test generation framework for C. ATLAS is open source and ships as a Docker image with a screencast walkthrough. Demo link: \url{https://youtu.be/50DvEbenp14}.

\end{abstract}

\section{Introduction}
\label{sec:introduction}

Software maintenance and machine learning for software engineering (ML4SE) tasks increasingly rely on source-level program views that expose syntax, control flow, and data flow. Multi-view code graphs align abstract syntax trees (ASTs)~\cite{ast-wiki}, control flow graphs (CFGs)~\cite{cfg-gfg}, and data flow graphs (DFGs)~\cite{dfg-sciencedirect} to capture syntactic structure, execution paths, and variable dependencies. These graphs encode the syntactic structure, execution paths, and variable dependencies that text-based models cannot observe directly. They support tasks such as test generation, code search, and vulnerability detection~\cite{10172747,liu2025toolindepthanalysiscode,10.5555/1855741.1855756,fontes2023integration,8638573}. Works like GraphCodeBERT~\cite{guo2020graphcodebert} and CodeSAM~\cite{mathai2024codesamsourcecoderepresentation} show that models using structural views consistently outperform text-based baselines. Yet for C/C++, to our knowledge, no single existing tool produces these aligned graphs without a complete build environment.

Most available tools address only parts of this problem. They are (a) restricted to compiled code and operate on intermediate representations (IRs) that discard source-level variable names~\cite{schubert2019phasar}, (b) limited to fixed non-configurable graphs or syntax parsing with no control or data flow~\cite{joern,tree-sitter}, and (c) designed for memory-managed languages only~\cite{das2023comex}. Extending existing multi-view frameworks to C and C++ requires reimplementing core analysis, because pointer aliasing, manual memory management, and the absence of runtime type information (RTTI) each break assumptions those frameworks rely on~\cite{schubert_et_al:LIPIcs.ECOOP.2021.2}. Practitioners must therefore combine several single-purpose tools to obtain all three views.

We present ATLAS, a command-line tool for multi-view code representation of C and C++ that takes a single source file or a multi-file project and emits an AST, a source-level inter-procedural CFG, a reaching-definition DFG, or any combination, in JSON, DOT, or PNG. It models data flow through references and aggregate members without whole-program points-to analysis. We make the following contributions.

\begin{itemize}
\item Build-free extraction that runs on partial inputs such as a submodule with missing headers, without requiring a build or compilation database.
\item Cross-view node alignment through shared identifiers, which lets users recover any single view by filtering edges, with no re-analysis.
\item Three CLI flags, view selection, variable collapsing, and node blacklisting, that resize the emitted graph without re-running the analysis.
\item Empirical evidence of utility on real C and C++ corpora and as the input to a downstream test-generation system.
\end{itemize}

On the TheAlgorithms corpora, ATLAS produces a correct CFG for $96.80\%$ of C files and $91.67\%$ of C++ files, including files that cannot be compiled due to missing dependencies. As downstream evidence, ATLAS has been adopted as the CFG front-end of SPARC~\cite{chowdhury2026sparcscenarioplanningreasoning}, an LLM-based unit-test generation framework in C, where it raises test coverage over a no-context baseline (Section~\ref{sec:evaluation:downstream}). ATLAS lowers the barrier for ML4SE researchers who need aligned multi-view graphs for C/C++.

\section{Related Tools}
\label{sec:related_works}

\begin{figure}[t]
  \centering
  \includegraphics[width=\columnwidth]{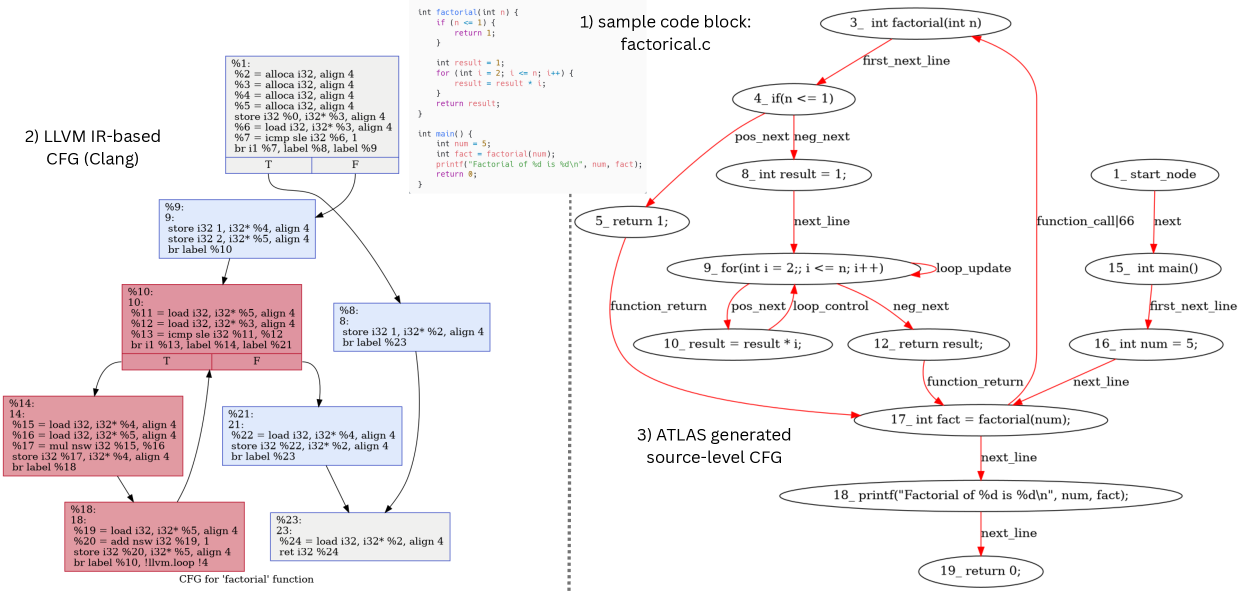}
  \caption{Comparison of CFG representations for the same C program: Clang/LLVM IR (left) versus ATLAS source-level CFG (right).}
  \label{fig:cfg_comparison}
\end{figure}

Compiler-based frameworks such as SVF~\cite{10.1145/2892208.2892235} and PhASAR~\cite{schubert2019phasar}, which operate on LLVM~\cite{lattner2004llvm} IR, provide precise analysis but require a resolved build environment. They also operate on IRs that discard source-level variable names and structural context, which downstream models need in order to map results back to code (Figure~\ref{fig:cfg_comparison}). Query-based analyzers such as CodeQL~\cite{avgustinov_et_al:LIPIcs.ECOOP.2016.2} and the Clang Static Analyzer~\cite{clangsa} share the same build requirement.

In contrast, Joern~\cite{joern} avoids the build requirement and supports C/C++. However, it provides no means to select which views to extract or to combine them into aligned multi-view representations. Its analysis is also primarily intra-procedural, so data flow that crosses function boundaries is not captured.

Similarly, Tree-sitter~\cite{tree-sitter} parses source code without compilation, but produces a syntax tree only. Without control or data flow, a model built on its output cannot reason about execution paths or variable dependencies, which are required for program analysis tasks.

Additionally, COMEX~\cite{das2023comex} produces configurable multi-view graphs with shared node identifiers directly from source, but targets memory-managed languages only, and its core analysis relies on properties that do not hold in C/C++. Three C/C++ semantics break those properties. Pointer aliasing creates ambiguous data flow, manual memory management removes the safety guarantees that reachability analysis depends on, and the absence of RTTI leaves virtual and function-pointer call targets unresolved~\cite{schubert_et_al:LIPIcs.ECOOP.2021.2}. Extending an existing multi-view framework to C/C++ therefore requires rebuilding its core analysis rather than modifying it.

To our knowledge, ATLAS is the only tool that satisfies all three requirements. Table~\ref{tab:competitors} summarizes the comparison, where ATLAS is the only listed tool that combines source-level operation, inter-procedural CFG construction, partial-input tolerance, and shared-identifier multi-view alignment.

\begin{table}[t]
\caption{Feature Comparison with Related Multi-View and Static Analysis Tools for C/C++}
\label{tab:competitors}
\centering
\renewcommand{\arraystretch}{1.1}
\resizebox{\columnwidth}{!}{%
\begin{tabular}{l|c|c|c|c|c|c}
\hline
\textbf{Tool} & \textbf{Build req.} & \textbf{Source-level CFG} & \textbf{Inter-proc. CFG} & \textbf{DFG} & \textbf{Multi-view aligned} & \textbf{Partial input} \\
\hline
ATLAS                                   & $\times$      & \checkmark & \checkmark & \checkmark & yes (shared ids) & \checkmark \\
Joern                                   & $\times$      & \checkmark & partial    & \checkmark & CPG (fixed)      & \checkmark \\
CodeQL                                  & \checkmark    & $\times$   & \checkmark & \checkmark & $\times$         & $\times$   \\
SVF / PhASAR                            & yes (LLVM IR) & $\times$   & \checkmark & \checkmark & $\times$         & $\times$   \\
Clang Static Analyzer                   & \checkmark    & \checkmark & partial    & partial    & $\times$         & $\times$   \\
Tree-sitter                             & $\times$      & $\times$   & $\times$   & $\times$   & $\times$         & \checkmark \\
COMEX (Java / C\#)                      & \checkmark    & \checkmark & partial    & \checkmark & \checkmark       & $\times$   \\
\hline
\end{tabular}%
}
\end{table}

\section{The ATLAS Tool}
\label{sec:atlas}

ATLAS is a command-line interface (CLI) that takes one or more C/C++ source files and emits a set of aligned graph representations, working on source-level syntax trees rather than a compiler IR. Its parser accepts any syntactically valid C/C++ regardless of whether the referenced types, functions, or templates are defined in the analyzed files, so it can analyze extracted submodules, auto-generated code, and incomplete codebases that do not build.

\begin{figure}[t]
  \centering
  \includegraphics[width=0.85\columnwidth]{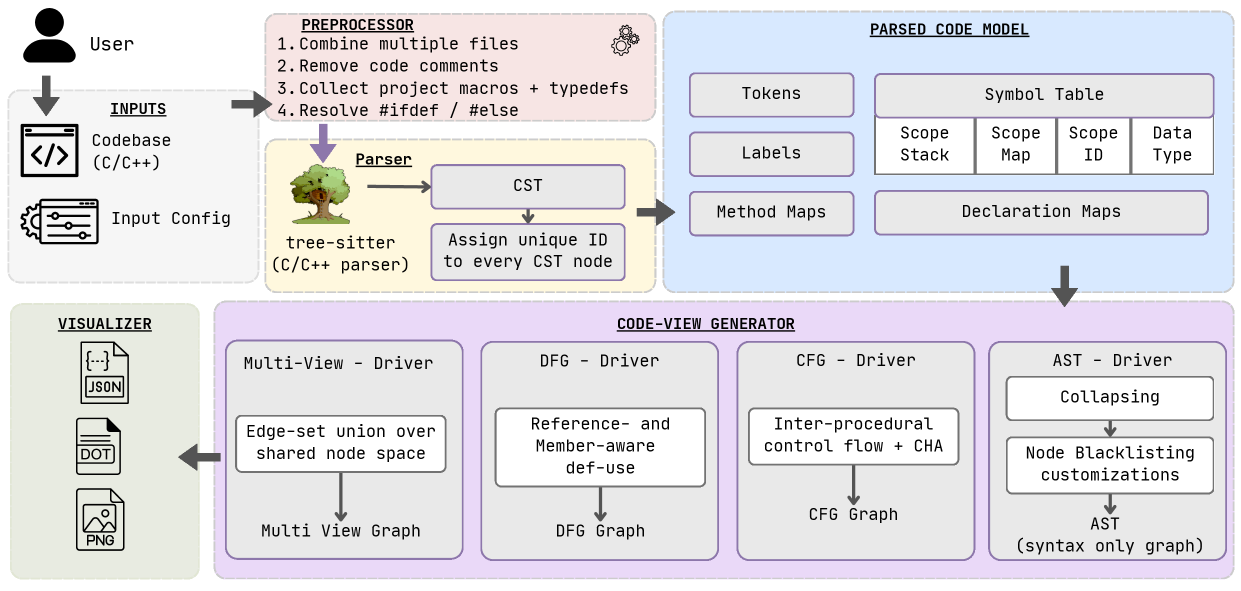}
  \caption{Overview of the \textsc{Atlas} four-stage pipeline: Preprocessor, Code-View Generator (AST, CFG, DFG, Multi-View drivers), and Visualizer.}
  \label{fig:overview}
\end{figure}

ATLAS runs a four-stage pipeline (Figure~\ref{fig:overview}): preprocessor, parser, code-view generator, and visualizer. It is invoked in one command, e.g.\ \texttt{atlas --lang c|cpp --code-folder PATH --graphs ast,cfg,dfg --output all}.

\subsection{System Architecture}
\label{sec:atlas:architecture}

\textbf{Preprocessor.} The preprocessor merges a multi-file project into one file, collecting project-local macros and \texttt{typedef}s and inlining intra-project headers while keeping system headers as opaque declarations. For conditional compilation, C keeps both \texttt{\#ifdef}/\texttt{\#else} arms and C++ takes the arm selected by in-source \texttt{\#define} values.

\textbf{Parser.} The parser runs Tree-sitter~\cite{tree-sitter} over the merged file to build a syntax tree, a symbol table, and a unique integer identifier for every node. Symbols of unknown type are tagged \texttt{unknown\_t} and treated as wildcards, so they do not block call-graph and def--use resolution. Because all drivers add edges over the shared identifiers, the AST, CFG, and DFG share one node namespace.

\textbf{Code-View Generator.} Four drivers (AST, CFG, DFG, Multi-View) operate over this model in the shared identifier space. Two options apply to every driver. Variable collapsing merges repeated identifiers into one node, and node blacklisting drops all nodes of a chosen syntactic category. Both reduce graph size before training.

\textbf{Visualizer.} Each view is exported as JSON for machines, Graphviz~\cite{gansner2000graphviz} DOT for other tools, and PNG for visual inspection. Multiple requested views are merged into one graph with edges colored by view class.

\subsection{Code Views}
\label{sec:atlas:views}

\begin{figure*}[t!]
    \centering
    \includegraphics[width=\textwidth]{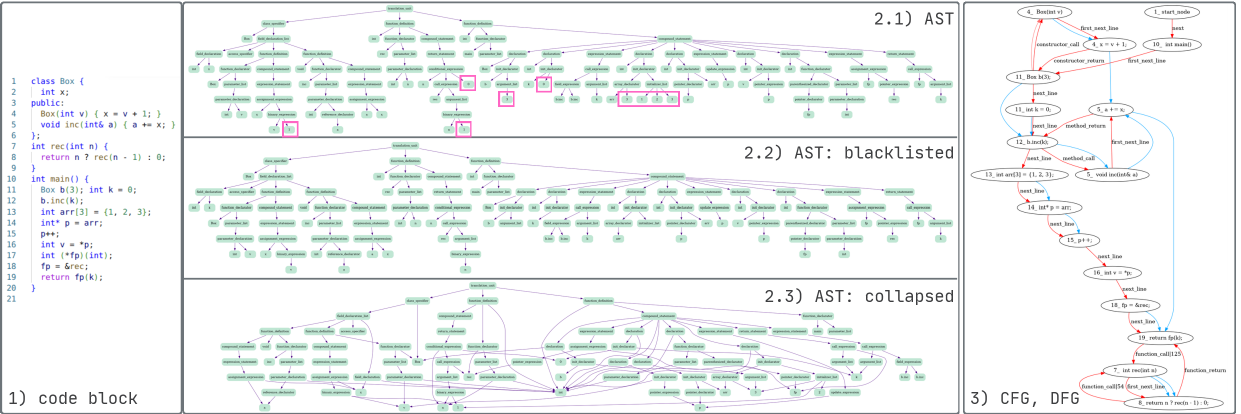}
    \caption{Visual representations of code analysis structures by ATLAS:
    \textbf{1)} Sample C++ code block.
    \textbf{2.1)} Default AST showing the full hierarchical syntactic structure of the source code (123 nodes, 122 edges).
    \textbf{2.2)} Blacklisted AST with numeric literals removed, reducing noise while preserving structural and identifier nodes (114 nodes, 113 edges). Highlighted nodes in (2.1) are the numeric literals absent from (2.2).
    \textbf{2.3)} Collapsed AST with repeated occurrences of the same identifier merged into a single node, reducing redundancy (89 nodes, 122 edges).
    \textbf{3)} Combined CFG+DFG overlaying labeled control-flow edges (red) and def--use data-flow edges (blue) on a shared set of source-statement nodes.}
    \label{fig:code_representations}
\end{figure*}

\textbf{AST.} The AST driver prunes the Concrete Syntax Tree (CST) of nodes with no semantic content, leaving a syntax-only graph whose nodes carry a grammar category and source range. It is the view on which variable collapsing and node blacklisting are most often applied. Figure~\ref{fig:code_representations}(2.1--2.3) shows their effect on the running example. The default AST has 123 nodes (2.1). Passing \texttt{--blacklisted number\_literal} prunes the numeric-literal leaves to 114 nodes (2.2), and \texttt{--collapsed} merges each repeated variable into one node for 89 nodes (2.3).

\textbf{CFG.} The CFG driver builds a source statement-level, inter-procedural control-flow graph directly over the source, so each node keeps its source-level name and structure with no lowering or basic-block grouping. It resolves direct calls, virtual calls through Class Hierarchy Analysis (CHA)~\cite{10.1007/3-540-49538-X_5}, and function-pointer calls through the pointer's last visible assignment, fanning out to signature-compatible callees when a target is unresolved (Section~\ref{sec:evaluation:callgraph}).

\textbf{DFG.} The DFG driver adds reaching-definition~\cite{10.1145/512927.512945} edges at the same statement granularity, with two inter-procedural flows that plain def--use tracking would miss. A pass-by-reference argument shares its parameter's definition so callee modifications reach the caller, and a member access keeps the object's identity across calls. Aliasing through pointer arithmetic, heap-stored pointers, or \texttt{void*} callbacks is not modeled, since ATLAS has no build and thus no points-to graph.

\textbf{Multi-View.} Because identifiers are shared, a multi-view output is just the union of the selected views' edges over one node set, for any non-empty subset of \{AST, CFG, DFG\}. Figure~\ref{fig:code_representations}(3) shows the combined CFG and DFG for the running example, with control-flow edges in red and def--use edges in blue over one shared set of source-statement nodes.

\section{Demonstration}
\label{sec:usage_demonstration}

\subsection{Installation and Invocation}
\label{sec:usage:install}
ATLAS ships as a Docker image. A single command builds it, and no system-wide toolchain or build database is required.
\begin{verbatim}
$ docker build -t atlas .
$ atlas --lang c --code-folder PATH \
        --graphs ast,cfg,dfg --output all
\end{verbatim}

\begin{figure}[t]
  \centering
  \includegraphics[width=0.82\columnwidth]{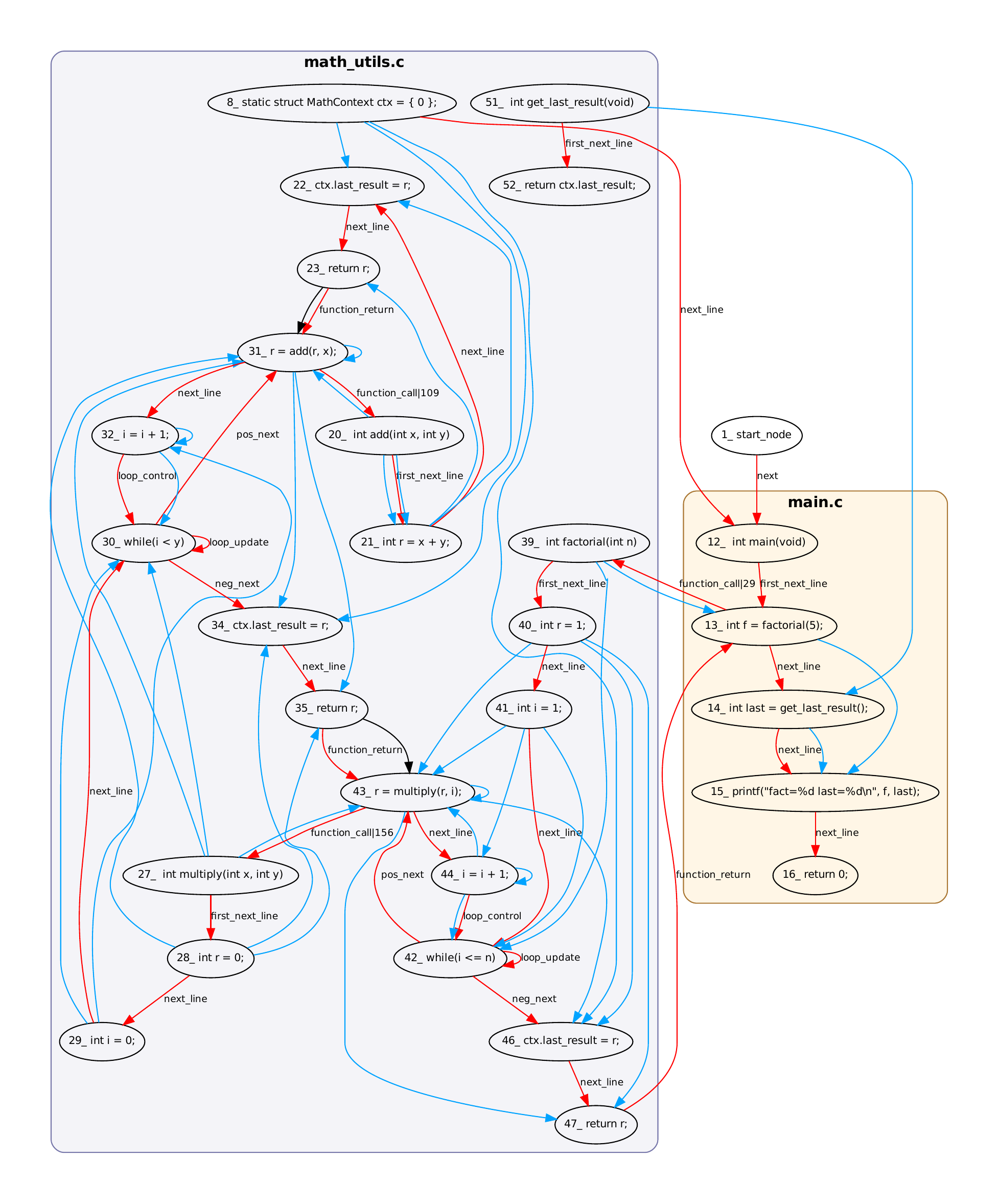}
  \caption{Combined CFG+DFG ATLAS produces for the three-file \texttt{math\_project} with \texttt{math\_utils.h} deleted. Red edges are inter-file calls (labeled \texttt{function\_call|\textit{lineno}}), blue edges inter-file def--use on the file-scope \texttt{ctx}. The missing header's \texttt{MathContext} type is tagged \texttt{unknown\_t} but still does not block edge construction (Section~\ref{sec:atlas:architecture}).}
  \label{fig:scenario_cfg}
\end{figure}

\subsection{Scenario: Analyzing a Project with a Missing Header}
\label{sec:usage:scenario}
This section walks through a single end-to-end use of ATLAS on a multi-file C project to show that ATLAS produces a connected inter-procedural CFG and DFG \emph{across source files} even when the header declaring the shared interface is missing from disk.
The header \texttt{math\_utils.h} declares the functions \texttt{add}, \texttt{multiply}, \texttt{factorial}, and \texttt{get\_last\_result}, together with a small struct \texttt{MathContext \{ int last\_result; \}}. The \texttt{math\_utils.c} implements the four functions and writes each computed result to a file-scope \texttt{static struct MathContext ctx}. \texttt{main.c} calls \texttt{factorial(5)}, then reads \texttt{ctx.last\_result} indirectly through \texttt{get\_last\_result()}. We remove \texttt{math\_utils.h} from disk before invocation.

The input is the three-file project \texttt{math\_project/}, comprising \texttt{math\_utils.h} (deleted), \texttt{math\_utils.c} (35 lines), and \texttt{main.c} (8 lines). A compiler-based pipeline cannot proceed on this input: the \texttt{\#include "math\_utils.h"} directive on line 1 of \texttt{math\_utils.c} fails to locate the header, and parsing halts before any analysis can run. ATLAS proceeds because its parser admits any syntactically-valid C without resolving \texttt{\#include}d declarations. The type \texttt{MathContext} referenced by the file-scope \texttt{ctx} declaration is recorded in the symbol table with an \texttt{unknown\_t} tag (Section~\ref{sec:atlas:architecture}), which downstream drivers treat as a wildcard at signature-matching and field-access sites:

\begin{verbatim}
$ clang -c math_utils.c
math_utils.c:1:10: fatal error:
    'math_utils.h' file not found

$ atlas --lang c \
        --code-folder math_project \
        --combined-name math_analysis \
        --graphs cfg,dfg --output all
[ok] output/math_analysis.{json,dot,png}
\end{verbatim}

Figure~\ref{fig:scenario_cfg} shows the resulting combined CFG+DFG. ATLAS itself emits a single merged graph. We group nodes into per-file clusters as a post-processing step on the DOT output. The inter-file call edge from \texttt{main}'s \texttt{int f = factorial(5);} into \texttt{factorial}'s entry node is constructed despite the missing declaration as signature matching against the visible definition in \texttt{math\_utils.c} is sufficient. The inter-file def--use chain on the file-scope \texttt{ctx} variable is constructed for the same reason as the field \texttt{ctx.last\_result} is written inside \texttt{add}, \texttt{multiply}, and \texttt{factorial}, and read from \texttt{get\_last\_result()}, which \texttt{main} calls. Each of these edges crosses the cluster boundary in Figure~\ref{fig:scenario_cfg}, and each carries the same node identifier scheme described in Section~\ref{sec:atlas:architecture}. ATLAS also emits the same graph as node-link JSON.

Restoring \texttt{math\_utils.h} and re-running ATLAS produces a structurally-equivalent graph, differing only by the single \texttt{declaration} node the header itself contributes (29 nodes / 74 edges without it, 30 nodes / 75 edges with). The partial-input graph is therefore a subset of the full graph, not an approximation of it. The same \texttt{--collapsed} and \texttt{--blacklisted} flags resize the emitted graph without re-running the analysis, letting a downstream model trade structural detail for graph size (Figure~\ref{fig:code_representations}).

\section{Evaluation}
\label{sec:evaluation}

We evaluate ATLAS along three dimensions. First, the fraction of a representative C and C++ corpus on which it produces correct output. Second, the scope of its inter-procedural call resolution and its runtime cost on production codebases in both languages. Third, the utility of its output as input to a downstream ML4SE task.

\subsection{Coverage of Language Constructs}
\label{sec:evaluation:coverage}

We applied ATLAS to the \texttt{TheAlgorithms} C and C++ datasets~\cite{TheAlgorithms}, which contain 406 C and 360 C++ files. ATLAS produces a non-empty CFG and DFG for every syntactically valid input. To validate correctness, we fixed the property checklist before any inspection began, then had three of the authors independently check each emitted graph against its source. A graph counts as correct only when it passes every property on the checklist, namely matching function entries, branch and loop topology, switch fan-out, return edges, same-unit inter-procedural calls, and reaching-definition consistency for unambiguous def--use pairs. We reconcile the three independent judgments by majority vote. ATLAS produces a correct CFG for 393 C files ($96.80\%$) and 330 C++ files ($91.67\%$). It produces a correct DFG for 371 C files ($91.38\%$) and 326 C++ files ($90.56\%$). Each remaining file contains exactly one of five constructs the current implementation does not model (Table~\ref{tab:unsupported}). These are documented limitations of the support matrix, not implementation bugs.

\begingroup
\tiny
\begin{table}[h]
    \setlength{\tabcolsep}{3pt}
\caption{Unsupported Constructs in the \texttt{TheAlgorithms} Corpus.
CFG failures are a subset of DFG failures, since the DFG depends on
correctness of the CFG. The per-language totals therefore count distinct files (35 in C,
34 in C++) rather than the sum of the CFG and DFG columns.}
\label{tab:unsupported}
\centering
\renewcommand{\arraystretch}{1.1}
\begin{tabular}{l|cc|cc}
\hline
& \multicolumn{2}{c|}{\textbf{C} (n=406)} & \multicolumn{2}{c}{\textbf{C++} (n=360)} \\
\textbf{Construct} & \textbf{CFG} & \textbf{DFG} & \textbf{CFG} & \textbf{DFG} \\
\hline
\texttt{goto} statements           & 3  & 3  & 0  & 0  \\
Multi-threading                    & 10 & 10 & 6  & 6  \\
Runtime pointer arithmetic         & 0  & 22 & 0  & 2  \\
Operator overloading               & 0  & 0  & 24 & 24 \\
Static class members               & 0  & 0  & 0  & 2  \\
\hline
\textbf{Total unsupported}         & \textbf{13} & \textbf{35} & \textbf{30} & \textbf{34} \\
\hline
\end{tabular}
\end{table}
\endgroup

\subsection{Resolution and Runtime Cost on Production Codebases}
\label{sec:evaluation:callgraph}

We ran ATLAS on two production codebases chosen for heavy indirect dispatch. On LevelDB~\cite{leveldb} (C++, roughly 10K~LOC), it finds 389 virtual and indirect call sites, and 92\% of them resolve to a single callee. On libcurl~\cite{curl} (C, roughly 124K~LOC), every same-unit direct call resolves correctly, and only 1.58\% of call sites are left unconnected, the ones that dispatch through struct-stored function pointers (Section~\ref{sec:limitations}).

On the 20 largest C programs in the corpus (61 to 309 LOC), CFG generation takes a median of 6.89~s at 135~MB peak memory on a single workstation (i7-10700K, 32GB DDR4 RAM, Ubuntu 24.04 LTS). The full libcurl run completes in roughly 23~minutes at about 534~MB.

\subsection{Downstream Use}
\label{sec:evaluation:downstream}

ATLAS-generated CFGs serve as the structural front-end of SPARC~\cite{chowdhury2026sparcscenarioplanningreasoning}, an LLM-based unit-test generation framework for automated unit-test generation in C. SPARC consumes source-level control-flow paths directly from ATLAS's CFG JSON output. On 59 real-world and TheAlgorithms C subjects it reports a $31.36\%$ absolute gain in line coverage and $26.01\%$ in branch coverage over a no-context prompting baseline, and matches or exceeds KLEE~\cite{10.5555/1855741.1855756} on its more complex subjects (full results in~\cite{chowdhury2026sparcscenarioplanningreasoning}). As a result, ATLAS's output carries enough structural context for a downstream LLM-based consumer to use it directly.

\section{Limitations and Future Work}
\label{sec:limitations}

ATLAS performs no points-to analysis, so the DFG misses aliasing through pointer arithmetic, heap-stored pointers, and \texttt{void*} callbacks, resulting in the 1.58\% of unconnected libcurl call sites in Section~\ref{sec:evaluation:callgraph}. An Andersen-style analysis would close most of this gap and is on the roadmap. Without a full C++ frontend, ATLAS approximates templates, exceptions, and complex inheritance. Operator overloading, \texttt{goto}, and concurrency are out of scope (Table~\ref{tab:unsupported}). We have validated graph correctness only on small files (Section~\ref{sec:evaluation:coverage}), not at the $10^{5}$-LOC scale. A large-project sweep and a user study of real-world adoption are our main future work.

\section{Availability and Reproducibility}
\label{sec:tool_availability}

ATLAS is released under the MIT License at \url{https://github.com/jaid-monwar/ATLAS-multi-view-code-representation-tool} which includes the Docker image, tool usage instructions, and paper artifacts. A 5-minute screencast is available at \url{https://youtu.be/50DvEbenp14}.

\bibliographystyle{IEEEtran}
\bibliography{references}

\end{document}